\documentclass{cimento1}
\usepackage{amsmath}

\title{QCD+QED lattice calculation of hadronic decay rates \thanks{{\it Presented at ``XVII edition of Incontri di Fisica delle Alte Energie'', Milano (Italy), 04--06 April 2018.}}}
\author{D.~Giusti (on behalf of the Soton--RM123 Collaboration)}
\instlist{
\inst
Dipartimento di Matematica e Fisica, Universit\`a degli Studi Roma Tre - Roma, Italy
\inst
INFN, Sezione di Roma Tre - Roma, Italy
}

\begin{document}

\maketitle

\begin{abstract}

Isospin is an almost exact symmetry of strong interactions and the corrections to the isosymmetric limit are, in general, at the percent level. For several hadronic quantities relevant for flavour physics phenomenology, such as pseudoscalar meson masses or the kaon leptonic and semileptonic decay rates, these effects are of the same order of magnitude of the uncertainties quoted in current large-scale QCD simulations on the lattice and cannot be neglected anymore. In this contribution I discuss some recent results for the leptonic decay rates of light pseudoscalar mesons obtained by the Soton--RM123 Collaboration including the leading-order electromagnetic and strong isospin-breaking corrections in first principles lattice simulations. The adopted strategy is within the reach of present lattice technologies and it allows to determine electromagnetic corrections to physical observables for which delicate cancellations of infrared divergences occur in the intermediate steps of the calculation. The application of the developed method to the study of heavy-light meson and semileptonic decay rates is currently underway.

\end{abstract}

\section{Introduction}

In the past few years, using different methodologies, accurate lattice results including electromagnetic (e.m.) effects have been obtained for the hadron spectrum as in the case of the neutral-charged mass splittings of light pseudoscalar ($P$) mesons and baryons~\cite{ref:Giusti:2017dmp, ref:Borsanyi:2014jba}.
In this respect, I have presented some recent determinations of the pion, kaon and $D$-meson mass splittings computed by the RM123 Collaboration at the previous edition of the {\it IFAE} conference in Trieste (19--21 April 2017)~\cite{ref:Giusti:2017ncc}.

In refs.~\cite{ref:deDivitiis:2011eh, ref:deDivitiis:2013xla} the inclusion of isospin-breaking (IB) effects in lattice QCD simulations has been carried out developing a method, the RM123 approach, which consists in an expansion of the lattice path-integral in powers of the two small parameters $ \alpha_{em}$ and $\left( m_d - m_u \right)$, where $\alpha_{em} \approx \left( m_d - m_u \right) / \Lambda_{QCD} \approx 1 \%$.

While the calculation of e.m.~effects in the hadron spectrum does not suffer from infrared (IR) divergences, the same is not true in the case of hadronic amplitudes, where e.m.~IR divergences are present and cancel for well-defined, measurable physical quantities only after including diagrams containing both real and virtual photons~\cite{ref:Bloch:1937pw}. Thus, the presence of IR divergences requires the development of additional strategies to those used in the computation of e.m.~effects in the hadron spectrum. Such a new strategy was proposed in ref.~\cite{ref:Carrasco:2015xwa}, where the lattice determination of the decay rate of a charged $P$ meson into either a final $\ell^\pm \nu_\ell$ pair or $\ell^\pm \nu_\ell \gamma$ state was addressed.

The e.m.~corrections due to the exchange of a virtual photon depends on the structure of the decaying meson, since all momentum modes are involved, and therefore it must be computed non-perturbatively.
Instead the non-perturbative evaluation of the emission of a real photon is not strictly necessary~\cite{ref:Carrasco:2015xwa}.
Indeed, it is possible to compute the amplitude for the emission of a real photon in perturbation theory by limiting the maximum energy of the emitted photon in the meson rest frame, $\Delta E_\gamma$, to be small enough so that the internal structure of the decaying meson is not resolved, but is larger than the experimental energy resolution.

The IR divergences depend on the charge and the mass of the meson but not on its internal structure ({\it i.e.}, they are universal). Thus, they cancel between diagrams containing a virtual photon (computed non-perturbatively) and those with the emission of a real photon (calculated perturbatively).

Since the experimental rates $\Gamma (P_{\ell 2})$ are inclusive, structure-dependent contributions to the real photon emission should be included. According to the chiral perturbation theory (ChPT) predictions of ref.~\cite{ref:Cirigliano:2007ga}, however, these contributions are negligible in the case of both kaon and pion decays into muons, while the same does not hold as well in the case of final electrons (see ref.~\cite{ref:Carrasco:2015xwa}). This important finding will be investigated by an ongoing dedicated lattice study on the real photon emission amplitudes in light and heavy $P$-meson leptonic decays~\cite{ref:deDivitiis:2019uzm}.

\section{Leading IB corrections to charged pion and kaon decay rates into muons: results}

The inclusive decay rate $\Gamma (P^\pm \to \ell^\pm \nu_\ell \left[ \gamma \right])$ can be expressed as
 \begin{equation}
      \Gamma (P^\pm \to \ell^\pm \nu_\ell \left[ \gamma \right]) = \Gamma_P^{(tree)} \left( 1+ \delta R_P \right)~,
      \label{e.decay}
 \end{equation}
where $\Gamma_P^{(tree)}$ is the tree-level decay rate given by
\begin{equation}
      \Gamma_P^{(tree)} = \frac{G_F^2}{8 \pi} \vert V_{q_1 q_2} \vert^2 m_\ell^2 \left( 1- \frac{m_\ell^2}{M_P^2} \right)^2 \left[ f_P^{(0)}\right]^2 M_P
      \label{e.tree}
 \end{equation}
and $M_P$ is the physical mass of the charged $P$ meson, including both e.m.~and strong IB corrections. The superscript $(0)$ on a physical quantity denotes that it has been calculated in isosymmetric QCD (without QED). The $P$-meson decay constant, $f_P^{(0)}$ is defined by  $A_P^{(0)} \equiv \langle 0 \vert \bar q_2 \gamma_0 \gamma_5 \bar q_1 \vert P^{(0)} \rangle \equiv f_P^{(0)} M_P^{(0)}$ and the quantity $\delta R_P$ encodes the leading-order e.m.~and strong IB corrections to the tree-level decay rate.

After having extrapolated our lattice data to the physical pion mass and to the continuum and infinite volume limits, we have determined for the first time the IB corrections to the ratio of the inclusive decay rates of kaons and pions into muons~\cite{ref:Giusti:2017dwk}
\begin{equation}
     \delta R_{K \pi}^{phys} = -0.0122 (10)_{stat} (12)_{syst} = -0.0122 (16)~.
     \label{e.ratio}
\end{equation}
Our result~(\ref{e.ratio}) can be compared with the current estimate $\delta R_{K \pi}^{phys} = -0.0112 (21)$ from ref.~\cite{ref:Cirigliano:2011tm} adopted by the Particle Data Group (PDG)~\cite{ref:PDG}.
Using the experimental $K_{\mu 2}$ and $\pi_{\mu 2}$ decay rates ~\cite{ref:PDG} and adopting the $N_f = 2+1+1$ Flavour Lattice Averaging Group average $f_K^{(0)} / f_\pi^{(0)} = 1.1958 (26)$~\cite{ref:FLAG}, one gets
\begin{equation}
     \bigg \vert \frac{V_{us}}{V_{ud}} \bigg \vert = 0.23142 (24)_{exp} (54)_{th}~.
     \label{e.CKM}
\end{equation}
 
Our preliminary results for $\delta R_\pi$ and $\delta R_K$ at the physical pion mass, in the continuum and infinite-volume limits are
\begin{eqnarray}
        \delta R_\pi^{phys} & = & 0.0148 (16)_{stat} (21)_{syst} = 0.0148 (26)~, \label{e.result_pi} \\
        \delta R_K^{phys} & = & 0.0020 (6)_{stat} (19)_{syst} = 0.0020 (20)~. \label{e.result_K}
\end{eqnarray}
The systematic uncertainties of the above determinations are dominated by a conservative estimate ($\simeq 25\,\%$) of the impact of the ${\cal O} (\alpha_{em} \alpha_s^n)$, with $n \geq 1$, corrections to the renormalisation constants of quark bilinear and weak four-fermion operators. It is currently on the way~\cite{ref:DiCarlo:2019thl} an improved renormalisation procedure in which the bare lattice operators are renormalised non-perturbatively into the regularisation-independent momentum subtraction (RI$^\prime$-MOM) scheme including e.m.~corrections and subsequently matched perturbatively at ${\cal O} (\alpha_{em} \alpha_s (M_W))$ into the W-regularisation scheme appropriate for these calculations. Our findings~(\ref{e.result_pi})--(\ref{e.result_K}) can be compared with the ChPT predictions $\delta R_\pi^{phys} = 0.0176 (21)$ and $\delta R_K^{phys} = 0.0064 (24)$ obtained in ref.~\cite{ref:Cirigliano:2011tm} and adopted by the PDG~\cite{ref:PDG}. The difference is within one standard deviation in the case of $\delta R_\pi^{phys}$, while it is larger for $\delta R_K^{phys}$ reaching the level of $\sim 2$ standard deviations.
 
\acknowledgments
I warmly thank my colleagues of the Soton--RM123 Collaboration for the enjoyable and fruitful work on the topics covered in this contribution.

\end{document}